\documentclass [12pt]{article}
\usepackage{graphicx,amssymb,amsmath,bm}
\usepackage[cp1251]{inputenc}
\usepackage[english]{babel}
\usepackage{cite}

\makeatletter
\def\@biblabel#1{#1.\hskip-0.3em}
\makeatother

\begin{document}
\def\refname{\normalsize\centering{REFERENCES}}
\def\abstractname{}

\begin{center}
{\large\bf{Aggregation in dilute aqueous solutions of hydroxypropyl cellulose with salt ions}}
\end{center}

\begin{center}
Valery~I.~Kovalchuk
\end{center}

\begin{center}
\textit{Taras Shevchenko National University of Kyiv, Faculty of Physics \\
(64/13, Volodymyrska Str., Kyiv 01601, Ukraine)}
\end{center}

\begin{abstract}
Thermal behaviors of hydroxypropyl cellulose dilute aqueous solutions with impurities of Group~I alkali metal
ions (Li, Na, K, Rb and Cs chlorides) has been studied by means of static and dynamic light scattering measurements.
From the experimental data, it follows that at temperatures above the LCST and in the presence of salts, there
arise supramolecular associates (clusters) that are several times larger than the wavelength of visible light.
It is shown that the observed intensity of backscattered light is described by the tangent plane approximation
and the Mie scattering theory.

An approach is proposed to model the hydrophobic association of polymer molecules in solutions with salt ions.
A parameter driving the interfacial energy density at the polymer-solvent interface is introduced into the Cahn-Hilliard
equation, which describes the temporal evolution of the polymer concentration in a given point of space. In this way,
the presence of electrolytes in solution is modeled. It follows from the results of computer simulation that an increase
in the interfacial energy density (strengthening of hydrophobic interactions) leads to an increase in the size of
polymer aggregates.

\end{abstract}

\bigskip
\begin{center}
\bf 1. Introduction
\end{center}

The study of the solutions of cellulose derivatives is based on promising technologies in various fields, including the
food industry, construction, hydrocarbon production, aerospace materials, medicine and pharmaceuticals~\cite{Okano1998,
Kamide2005, Kabir2018}. A characteristic feature of many water-soluble cellulose ethers is the thermoreversible volume
transition~\cite{Tanaka2011}. Its essence consists in the formation of a polymer gel network when the temperature increases,
but the system returns to the state of isotropic solution when cooled down. The threshold temperature of this transition
is called the lower critical solution temperature (LCST) which depends on a number of factors such as polymer concentration,
type and degree of substitution~\cite{Bayer2012, Tian2019, Gosecki2021}, pH value~\cite{Zhang2011, Bai2012, Qiu2013, Ofridam2021},
and the presence of electrolytes in solution~\cite{Xu2004-1, Xu2004-2, Zheng2004, Liu2008, Weibenborn2019}.

A number of works are devoted to the study of the thermosensitive behavior of cellulose derivatives aqueous solutions with salt
ions~\cite{Nystrom1992, Xu2004-1, Xu2004-2, Zheng2004, Joshi2011, Fettaka2011, Almeida2014, Lazarenko2022, Lazarenko2023}.
The main methods that were used: turbidimetry~\cite{Zheng2004, Joshi2011, Fettaka2011, Lazarenko2023}, dynamic light scattering
(DLS)~\cite{Nystrom1992, Fettaka2011, Lazarenko2022, Lazarenko2023}, microcalorimetry~\cite{Xu2004-1, Xu2004-2, Zheng2004,
Joshi2011}, and viscometry~\cite{Xu2004-2, Zheng2004, Joshi2011, Almeida2014, Lazarenko2022, Lazarenko2023}. In the articles
listed above the volume phase transition mechanisms were mainly studied depending on the type of salt additives and their
concentration. In particular, it was shown that the thermosensitive behavior of polymer solution is based on competition for
water molecules between polymer chains and salt ions, which leads to the hydrophobic aggregates formation~\cite{Xu2004-1,
Xu2004-2, Zheng2004, Liu2008}.

It should be noted that ions can act as aggregation initiators not only in solutions of cellulose derivatives, but also in
other systems containing, for example, peptides~\cite{Klement2007}, pseudo-polypeptoids~\cite{Kirila2021}, lignin~\cite{Fritz2017},
lipid nanoparticles~\cite{Wei2011}, latex particles~\cite{Pefferkorn1992, Hanus2001}, and silica particles~\cite{Tavacoli2007}.
In general, the study of aggregation in polymer solutions is of great practical importance for modern technologies, such as
catalysts with controlled activity and film nanocomposite materials~\cite{Khokhlov1997, Hussain2015, Alharbi2023}, color-based
sensors that respond to pH and heavy metal ions~\cite{Lee2020}, energy-saving smart windows~\cite{Connelly2017, Nakamura2021},
and drug delivery systems~\cite{Zhang2011, Bai2012}.

Hydroxypropyl cellulose (HPC) as an object of research was chosen due to the attractiveness of its physical and chemical
properties. In addition to the well-known unique properties of a green polymer, such as availability, cheapness, and biocompatibility,
HPC has a low LCST value of \mbox{$(40\div45)^{\circ}$C}~\cite{Connelly2017, Nakamura2021, Zhang2011, Bai2012, Harsh1991, Narang2019,
Gosecki2021}. Such a temperature makes HPC a good choice for both biomedical applications and promising smart technologies~\cite{Qiu2013}.
In our recent work~\cite{Lazarenko2023}, it was found that the addition of salts (chlorides of Group~I alkali metal ions) to
dilute aqueous solutions of HPC leads to polymer aggregation at temperatures above LCST. As a continuation of~\cite{Lazarenko2023},
in this work these same systems are studied using static light scattering. The obtained light backscattering data were analyzed
within the framework of the tangent plane approximation and Mie scattering theory. It is also shown that the hydrophobic association
of polymer molecules in the presence of electrolytes in solution can be described within the framework of the Cahn-Hilliard model.


\bigskip
\begin{center}
\bf 2. Materials and Methods
\end{center}

{\flushleft
\textit{2.1. Materials}}

The cellulose derivative, hydroxypropyl cellulose (HPC), was purchased from Alfa Aesar company. The manufacturer's
specification indicates that HPC has an average degree of substitution of 75.7\% and a molecular weight
of 100,000. The viscosity range was reported by the manufacturer to be 112~cPs at $25^{\circ}$C for a
5~wt\% aqueous solution.

Analytical grade salts (Li, Na, K, Rb, Cs chlorides) were purchased from Sigma-Aldrich company.

{\flushleft
\textit{2.2. Specimen preparation}}

The initial aqueous HPC solution with a concentration of 2 wt\% was prepared by dissolving the necessary
amount of polymer in deionized water under continuous stirring for 4~hours at a temperature of $60^{\circ}$C
to ensure the complete dissolution of the polymer. As a result, a homogeneous and transparent solution was
obtained.

This initial solution was divided to prepare six specimens. The salts were introduced into five of them,
and the specimens were mixed until the complete salt dissolution. The molar concentrations of the salt in
the specimens were identical and equal to that of the physiological solution (154 mmol/l). By diluting 2\%
solutions with water to a ratio of 1:10, specimens with a polymer concentration of 0.2~wt\% and a salt
concentration of 15.4~mmol/l were fabricated.

{\flushleft \textit{2.3. Dynamic light scattering}}

Particle sizes (hydrodynamic diameters) were determined using a Zetasizer Nano ZS (Malvern, UK) instrument
at $173^{\circ}$ backscatter geometry. Before measurement, all specimens were dust-removed through a 0.45~um
filter (Minisart NML). For each temperature point, the particle size distributions in a given specimen were
measured three times.

{\flushleft
\textit{2.4. Static light scattering}}

Fig.~1 shows the schematic of the measuring chamber of the experimental setup used for studies of polymer
solutions by static light scattering~\cite{Kovalchuk2022, Kovalchuk2021}.

\begin{figure}[!h]
\center
\includegraphics [scale=0.5] {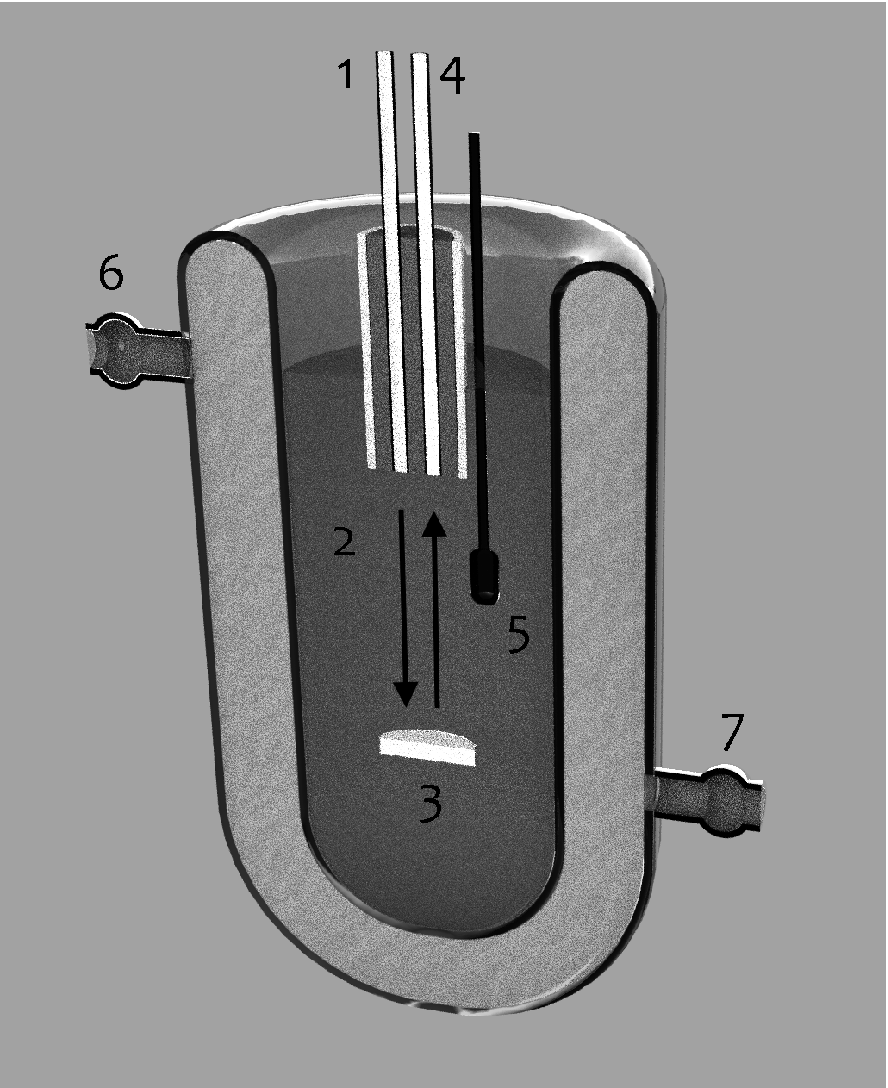}
\caption{Schematic of the measuring chamber of the experimental setup.}
\label{fig1}
\end{figure}

As light source, we used a GNL-5013PGC LED with a known spectral characteristic. It was powered using a
micropower current stabilizer made on an LP2951 chip. Light with a wavelength of 525~nm was fed through fiber
optic cable 1 into thermostatic chamber 2 filled with the polymer solution. The light beam reflected from
mirror 3 returned through optical fiber cable 4 to a digital photosensor connected to an ATmega328P microcontroller.
As a result, the total intensity
\begin{equation}
J_{\Sigma}=J_{\text{T}}+J_{\text{R}}
\label{eq1}
\end{equation}
was measured, where $J_{\text{T}}$ is the intensity of the light flux attenuated because of its passage through
the specimen, and $J_{\text{R}}$ is the intensity of the light flux scattered by the specimen at an angle of $180^{\circ}$.
When measuring $J_{\text{R}}$, mirror 3 was screened with curtain made of light-absorbing material (black anodized aluminum).
Thus, in order to determine $J_{\text{T}}$ and $J_{\text{R}}$ for a specific specimen, it was necessary to perform two
measurements: in the presence and in the absence of screen.

The solution temperature was measured by means of a digital temperature sensor LMT01LPG 5 and another an ATmega328P
microcontroller. The illumination and temperature values were synchronously read out using a microcontrollers and transferred
to USB ports of a personal computer using the RS-232 protocol. Accumulation of data, their further processing
and visualization were implemented with the help of a Delphi code. The illuminance measurement error did not exceed 0.1~lux,
and the temperature measurement error was $\pm0.1^{\circ}$. The chamber with the polymer solution was placed in light-proof
box and connected with the help of nozzles 6 and 7 to a circulation thermostat Julabo~ME-6. Structurally, elements 3 and 5
together with the output (input) of optical fiber cables 1 and 4 were arranged in the form of a probe that was immersed into
the examined solution.


\bigskip
\begin{center}
\bf 3. Results and Their Discussion
\end{center}

{\flushleft
\textit{3.1. Static and dynamic light scattering}}

When falling on the surface of the examined solution (see Fig.~1), the primary beam of light formed
two beams, transmitted and reflected. The corresponding intensities $J_{\text{T}}$ and $J_{\text{R}}$ of these beams
were measured at a set of temperatures within an interval from 30 to $65^{\circ}$C. The specimen heating rate was
\mbox{$1.1^{\circ}$}C/min. The obtained experimental data were calibrated in each specimen to a transparency level
of 100\%. As such the corresponding value of $J_{\text{T}}$ at a temperature of $30^{\circ}$C was selected.

Figure~2 exhibits the measurement results obtained for the temperature dependences of the relative intensities of
the backscattered light beam for all six specimens. The same figure also shows the average size of polymer particles
(hydrodynamic diameter, $D$) determined by the dynamic light scattering method (points)~\cite{Lazarenko2023}.

\begin{figure}[!h]
\center
\includegraphics [scale=0.95] {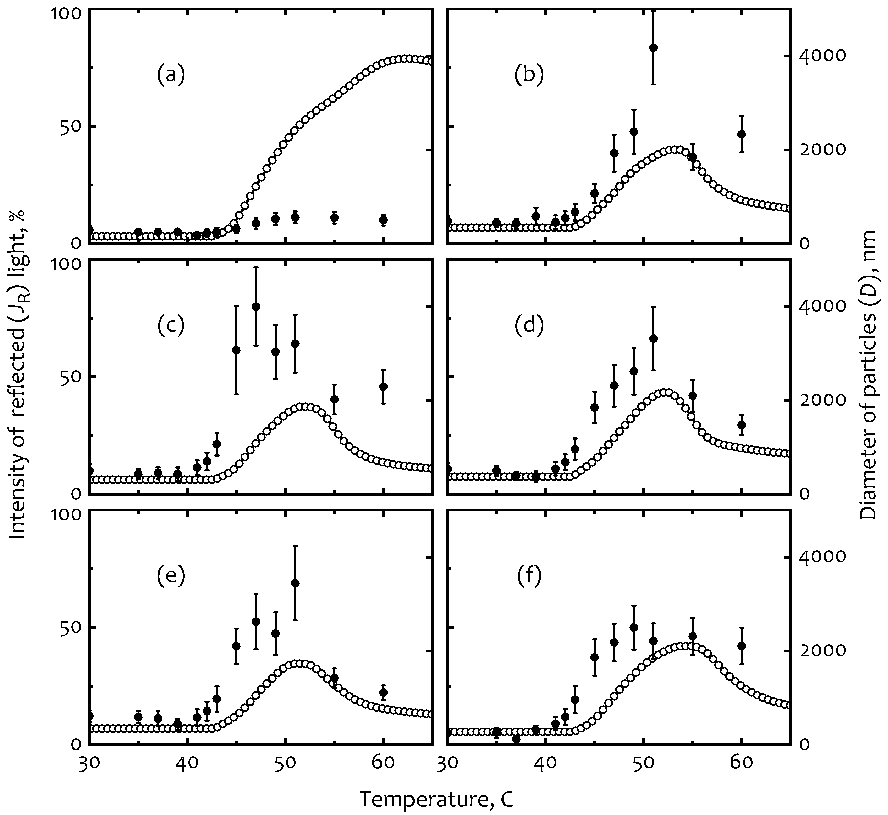}
\caption{Temperature dependences of the reflected light intensity $J_{\text{R}}$ (left vertical scale,~$\circ$)
for specimens: (a) -- ion-free, (b) -- Li, (c) -- Na, (d) -- K, (e) -- Rb, (f) -- Cs. The hydrodynamic diameter values
of the polymer clusters $D$ are also shown here (right vertical scale,~$\bullet$).}
\label{fig2}
\end{figure}

{\flushleft
\textit{3.2. Light Backscatter Analysis}}

Fig.~2 shows that the experimental dependences $J_{\text{R}}(T)$ for the solutions with salt ions have
a non-monotonic character: as the temperature increases, the $J_{\text{R}}$-values first grow, then reach
a maximum, and afterward decrease. This behavior can be explained as follows.

The solution surface on which the light falls is not perfectly smooth -- it contains polymer inclusions of
various sizes, therefore it can be considered as a statistically rough surface. The studies of light scattering by
such surfaces have a long (for more than a hundred years) story~\cite{Bass1979}. It is known that the solution of this
problem is reduced to the analysis of a wave equation of a certain type. Instead, in this paper, we will confine
ourselves to a qualitative consideration proceeding from assumptions concerning the type of backward scattering,
which forms the output beam with the intensity $J_{\text{R}}$.

Let the solution surface look like a plane on average. Let's draw the $z$-axis perpendicular to this plane.
Surface roughness \mbox{$z=\xi(r)$} is described by two statistical parameters~\cite{Lu2000}: parameter $\delta$
(root-mean-square height of surface deviation from the plane \mbox{$z=0$}) and parameter $\alpha$ (lateral
correlation length of roughness, describing the average lateral distance between the peak and valley of the
surface profile).

We assume that the incident light propagates along the $z$-axis and that the condition
\begin{equation}
(k R_{\text{c}})^{1/3}\gg{1}
\label{eq2}
\end{equation}
is satisfied. Here $k$ is the wave vector modulus of the photon, $R_{\text{c}}$ is the radius of surface curvature,
which can be estimated as \mbox{$R_{\text{c}}\simeq2\sqrt{3}\alpha^{2}/\delta$}~\cite{Lu1991}.

Inequality (\ref{eq2}) is a condition for the applicability of the tangent plane method, developed in~\cite{Voronovich2007}
based on the theory of wave scattering by a statistically rough surface~\cite{Bass1979}. The main points and conclusions
of this method as applied to backscattering of light are as follows.

When considering methods for assessing the reflective properties of light by a rough surface, the entire reflected
signal is divided into a coherent component and an incoherent one. The coherent component is associated with specular
reflection from surface areas and is determined by the average value of the field strength in the scattered wave.
According to~\cite{Bass1979}, the back reflection coefficient of the coherent component for a surface with a normal
distribution of heights is equal to
\begin{equation}
\rho = J_{\text{R}}/J_{0} = \exp(- 2k^{2}\delta^{2}),
\label{eq3}
\end{equation}
where $J_{0}$ is the intensity of incident light beam. From (\ref{eq3}) it follows that the value of $J_{\text{R}}$
decrease exponentially as the characteristic height of irregularities increases.

In~\cite{Bass1979, Kodis1966} it was shown that the cross section of backscattering of light by a three-dimensional
statistically rough surface is
\begin{equation}
\Sigma\sim{Na^{2}},
\label{eq4}
\end{equation}
where $N$ is the number of scattering centers (points of specular reflection, which correspond to the peaks and
valleys of the surface $z=\xi(r)$), and the value $a$ is the geometric mean of the main radii of curvature $a_{1}$
and $a_{2}$ at the points of specular reflection
\begin{equation}
a^{2} = \langle{a_{1}a_{2}}\rangle.
\label{eq5}
\end{equation}

Formula (\ref{eq4}) has a simple geometric meaning: the scattering cross section by a rough surface in the tangent
plane approximation (\ref{eq2}) coincides with the scattering cross section by $N$ identical balls with radius $a$.
Consequently, the problem of light scattering by a statistically rough surface can be reduced to the equivalent problem
of scattering by a system of spherical particles. Since the amount of polymer in the solution is a constant parameter,
the number of such particles equals
\begin{equation}
N \sim 1/a^{3},
\label{eq6}
\end{equation}
and, therefore,
\begin{equation}
\Sigma \sim 1/a.
\label{eq7}
\end{equation}

Note that the result (\ref{eq7}) can also be obtained in the framework of Mie scattering theory~\cite{Mie1908}.
The Mie backscattering cross section for a single particle has the form~\cite{Tzarouchis2018}
\begin{equation}
\sigma(a)=\frac{1}{a^{2}k^{2}}\Bigl| \sum_{n=1}^{\infty}(-1)^{n}(2n+1)
(A_{n}(a,k)-B_{n}(a,k)) \Bigr|^{2},
\label{eq8}
\end{equation}
where $A_{n}$, $B_{n}$ are the Mie coefficients\cite{Kerker1983}.

Figure~3 shows the computing results of $\sigma(a)$ that were performed using the~\cite{Matzler2002, Mie2024}
method for light with a wavelength of 525~nm and a refractive index value for 0.2\% water solution of HPC
taken from~\cite{Maklakova2019}.

\begin{figure}[!h]
\center
\includegraphics [scale=0.55] {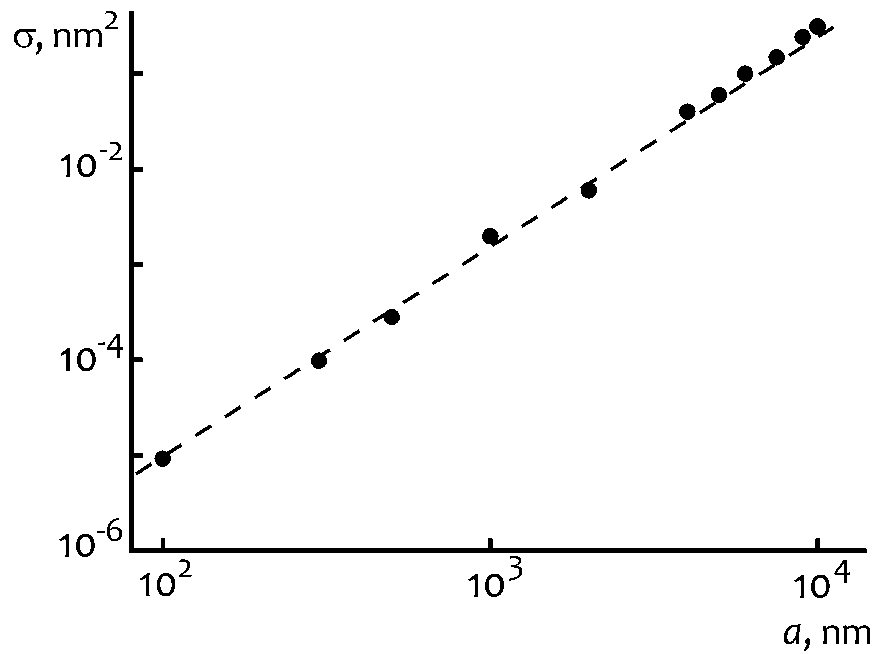}
\caption{Mie backscattering cross section for a single spherical aggregate of hydroxypropyl cellulose
with radius $a$ (set of black dots).}
\label{fig3}
\end{figure}

This figure shows that the $\sigma(a)$ dependence has the scaling behavior (dashed line)
\begin{equation}
\sigma(a)\sim{a^{s}}
\label{eq9}
\end{equation}
with the power exponent \mbox{$s=2$}.
Hence, the total cross section of Mie backscattering for $N$ identical spherical particles, taking into account
(\ref{eq6}), is
\begin{equation}
\Sigma_{\text{Mie}}\sim{Na^{2}} = 1/a,
\label{eq10}
\end{equation}
which coincides with the formula (\ref{eq7}). It follows from formula (\ref{eq10}) that
\begin{equation}
J_{\text{R}}\sim{1/a},
\label{eq11}
\end{equation}
that is, the intensity of backscattered light increases as the particle size decreases. The same conclusion can be
reached by comparing the experimental values of $J_{\text{R}}$ and $D$ in Figure~2a with the values of the same quantities
in any of the remaining Figures~2b-f.

As mentioned above, the behavior of $J_{\text{R}}(T)$ in Fig.~2b-f has a non-monotonic character: as the temperature
increases, the $J_{\text{R}}$-values first grow, then reach a maximum at \mbox{$T{\kern 1pt}'\simeq (50\div55)^{\circ}$}C,
and afterward decrease. The sizes of the polymer clusters (see the same figures) have a similar temperature behavior.
There are two explanations for the decrease of cluster sizes at \mbox{$T>T{\kern 1pt}'$}. The first of them consists
in that the clusters collapse at \mbox{$T>T{\kern 1pt}'$} and form numerous small fragments. But in this case,
as follows from Eq.(\ref{eq11}), instead of the reduction of the intensity $J_{\text{R}}$, we would observe its
growth, which contradicts the experiment. The second explanation is that large polymer aggregates disappear from
the solution at \mbox{$T>T{\kern 1pt}'$} owing to their sedimentation, which is confirmed by the experiment: whenever
the measurements were concluded, some polymer sediment was found at the bottom of the cell with the solution. This
sediment did not dissolve well in a hot solution, but it dissolved easily in cold water.

Note that the viscosity of the 0.2-wt\% aqueous solutions of HPC at \mbox{$(50\div60)^{\circ}$}C exceeds the viscosity
of water by only 5–7\%~\cite{Lazarenko2022}. Therefore, the sedimentation of polymer clusters occurred quite quickly:
from Fig.~2b-f, it follows that $J_{\text{R}}$ decreased by a factor of two during about 10~min.

Figure~2 also demonstrates that the sizes $D$ of clusters in the ion-free specimen are several times smaller than in
the specimen with ions. By order of magnitude, they are equal to the wavelength of visible light. In Fig.~2a, the intensity
of reflected light is larger than that in Fig.~2b-f, as it should be according to formula (\ref{eq11}). The polymer clusters
in the ion-free solution were in a suspended state, and practically no sediment was observed in the cell after the measurements.

{\flushleft
\textit{3.3. Computer simulation of HPC aggregation: the role of hydrophobic interactions}}

The hydrophobic association of the polymer in the vicinity of LCST can be investigated within the framework of the Cahn-Hilliard
model developed from the free energy density functional method.

The Ginzburg-Landau type functional for the total energy of the polymer-solvent system has the form~\cite{Li2015}
\begin{equation}
U[\phi] = \int d\vec{r}\,\Bigl\{F(\phi) + \kappa(\phi)|\nabla\phi|^{2} \Bigr\}, \quad
\phi = \phi(\vec{r},t),
\label{eq12}
\end{equation}
where $\phi$ is the order parameter meaning the polymer concentration in solution, $F(\phi)$ is the free energy density.
The second term in curly brackets under the integral sign in (\ref{eq12}) describes the contribution of spatial correlation
effects to the free energy with a gradient coefficient~\cite{DeGennes1980}
\begin{equation}
\kappa(\phi) = \frac{a^{2}}{36\phi(1-\phi)},
\label{eq13}
\end{equation}
where $a$ is the size of the polymer chain segments, equal to 1 for the Flory model~\cite{Flory1953}.

Let's introduce an additional parameter $\gamma$ into (\ref{eq12}):
\begin{equation}
U[\phi] = \int d\vec{r}\,\Bigl\{F(\phi) + \gamma\kappa(\phi)|\nabla\phi|^{2} \Bigr\},
\label{eq14}
\end{equation}
the physical meaning of which is as follows.

In the works of Cahn and Hilliard~\cite{Cahn1958, Cahn1961} it was shown that the interfacial surface tension coefficient
\begin{equation}
\sigma_{\text{i}} \sim \int d\vec{r}\,\kappa(\phi)|\nabla\phi|^{2}.
\label{eq15}
\end{equation}
Macromolecules of cellulose derivatives are amphiphilic, i.e. they contain both nonpolar hydrophobic and polar hydrophilic
sites in their structure. The latter compete with salt ions for water molecules in solutions with electrolytes~\cite{Fettaka2011,
Xu2004-1, Xu2004-2, Zheng2004}. As a result, the number of hydrophilic water-polymer bonds decreases and the interfacial
energy density, respectively, increases. In other words, the enhancement of hydrophobic interactions at the polymer-solvent
interface leads to an increase in $\sigma_{\text{i}}$. Thus, by changing the value of the parameter $\gamma$ in the formula
(\ref{eq14}), the hydrophobicity of the polymer can be simulated.

The Cahn-Hilliard equation describing the evolution of the polymer concentration at a point in space $\vec{r}$ at time $t$
has the form~\cite{Li2015}
\begin{equation}
\frac{\partial\phi}{\partial{t}} = \nabla
\Bigl\{
M\nabla\frac{\delta{U[\phi]}}{\delta\phi}
\Bigr\} + \xi,
\label{eq16}
\end{equation}
where $M$ is the mobility, \mbox{$\xi=\xi(\vec{r},t)$} is a stochastic function (thermal noise) satisfying the
fluctuation-dissipative theorem~\cite{Landau1980}. To solve equation (\ref{eq16}) with the functional (\ref{eq14}),
we use the Flory-Huggins free energy~\cite{Flory1953, Huggins2013} in the form
\begin{equation}
F(\phi) = N^{-1}\phi\ln\phi + (1-\phi)\ln(1-\phi) + \chi\phi(1-\phi),
\label{eq17}
\end{equation}
where $N$ is the degree of polymerization of the molecular chain, $\chi$ is the Flory-Huggins parameter describing pairwise
interactions between monomers. Then the variational derivative of the total energy in (\ref{eq16}) is equal to
\begin{equation}
\frac{\delta{U[\phi]}}{\delta\phi} = f(\phi) -
\gamma
\bigl(
\lambda(\phi)|\nabla\phi|^{2} + 2\kappa(\phi)\Delta\phi
\bigr),
\label{eq18}
\end{equation}
where
\begin{equation}
f(\phi) = N^{-1}(1+\ln\phi) - \ln(1-\phi) - \chi(2\phi-1) - 1,
\label{eq19}
\end{equation}
\begin{equation}
\lambda(\phi) = \frac{2\phi-1}{36\phi^{2}(1-\phi)^{2}}.
\label{eq20}
\end{equation}
We consider that the mobility of polymer molecules depends on their concentration
\begin{equation}
M = M_{0}\phi(1-\phi).
\label{eq21}
\end{equation}
Finally, the Cahn-Hilliard equation takes the form
\begin{equation}
\frac{\partial\phi}{\partial{t}} =
M_{0}\nabla
\Bigl\{
\phi(1-\phi)\nabla
\bigl(
f(\phi)-
\gamma
(
\lambda(\phi)|\nabla\phi|^{2} + 2\kappa(\phi)\Delta\phi
)
\bigr)
\Bigr\}
+\xi.
\label{eq22}
\end{equation}

This is a fourth-order nonlinear parabolic partial differential equation with a stochastic term. To enforce
the mass conservation law and descending of total energy with time, equation (\ref{eq22}) must be equipped
with the homogeneous Neumann boundary conditions~\cite{Lee2014}.

Without loss of generality, let us investigate solutions $\phi(x,t)$ of the one-dimensional equation (\ref{eq22}).
Introducing a spatio-temporal grid for $x$ and $t$ with periodic boundary conditions, we use for solving of
(\ref{eq22}) the semi-implicit difference scheme~\cite{Li2015} with the parameters of the polymer-solvent system
under study taken from~\cite{Kovalchuk2021, Orwoll2007}. The initial phase was homogeneous, and the corresponding
polymer concentration for a 0.2~wt\% solution was $\phi_{\text{in}}=2\cdot10^{-3}\rho$, where \mbox{$\rho=1.3\,\,
\text{g/cm}^{3}$} is the density of hydroxypropyl cellulose~\cite{HPC}. The phase structure growth was simulated
on a lattice with $L=128$.

In Figure~4, as an example, the result of numerical simulation is demonstrated for \mbox{$\gamma=0.04$}, where
it is seen how the reduced concentration \mbox{$\overline{\phi}=\phi/\phi_{\text{in}}$} changes depending on the
dimensionless distance $x$ and time $t$. It can be seen how an unstable regime arises on the basis of a random
distribution \mbox{$(t=0.001)$}, which leads to the primary structure of a new phase \mbox{$(t=0.2)$}, and then
to its further coarsening and final formation \mbox{$(t=1-5)$}.

\begin{figure}[!h] 
\centering
\includegraphics [scale=0.9] {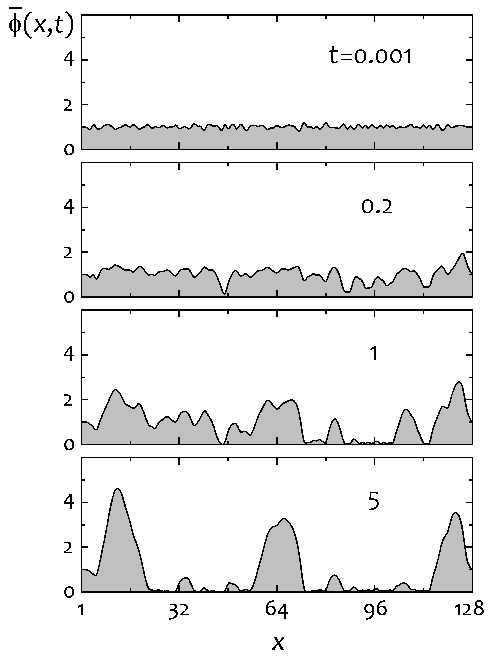}
\caption{Distribution of the reduced concentration \mbox{$\overline{\phi}(x,t)$} on a one-dimensional lattice during
the spinodal decomposition in the hydroxypropyl cellulose-water system. The value of \mbox{$\gamma$} is 0.04.}
\label{fig4}
\end{figure}

One of the features of the asymptotic behavior of the system under study at large times is dynamical scaling,
which is determined by the characteristic length $L(t)$ of the individual ordered regions of the phase
structure~\cite{Bray1994}. Choosing the average radius of cluster $R(t)$ as characteristic length at the level
\mbox{$\overline{\phi}=1$}, we find the scaling exponent for the growth law \mbox{$R(t)\sim{t^{\delta}}$}.
Analysis of the time dependencies $R(t)$ in their longest linear sections in Figure~5 gives the value
\mbox{$\delta=0.324\pm{0.026}$}. This value is close to the exponent \mbox{$\delta=1/3$} which is typical
for the laws of growth in systems with a conserved scalar order parameter~\cite{Lifshitz1961,Bray1993}.

\begin{figure}[!h] 
\centering
\includegraphics [scale=0.5] {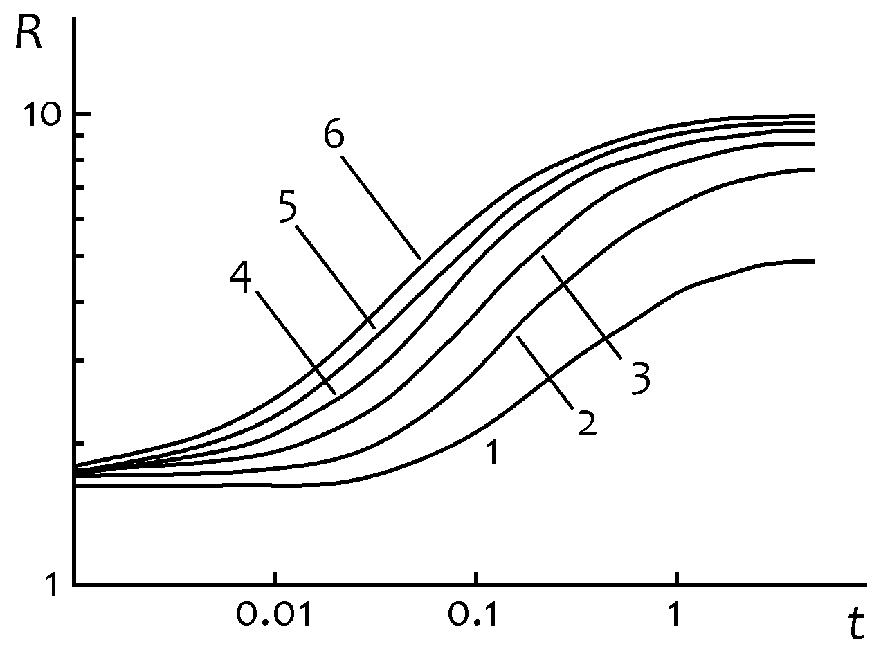}
\caption{Dependencies of the average cluster radius over time for the set of $\gamma$ values: 0.01~(1),
0.02~(2), 0.03~(3), 0.04~(4), 0.05~(5), 0.06~(6). Averaging was performed for 100 runs of the program
simulating the growth of the phase structure.}
\label{fig5}
\end{figure}

\begin{figure}[!h] 
\centering
\includegraphics [scale=0.5] {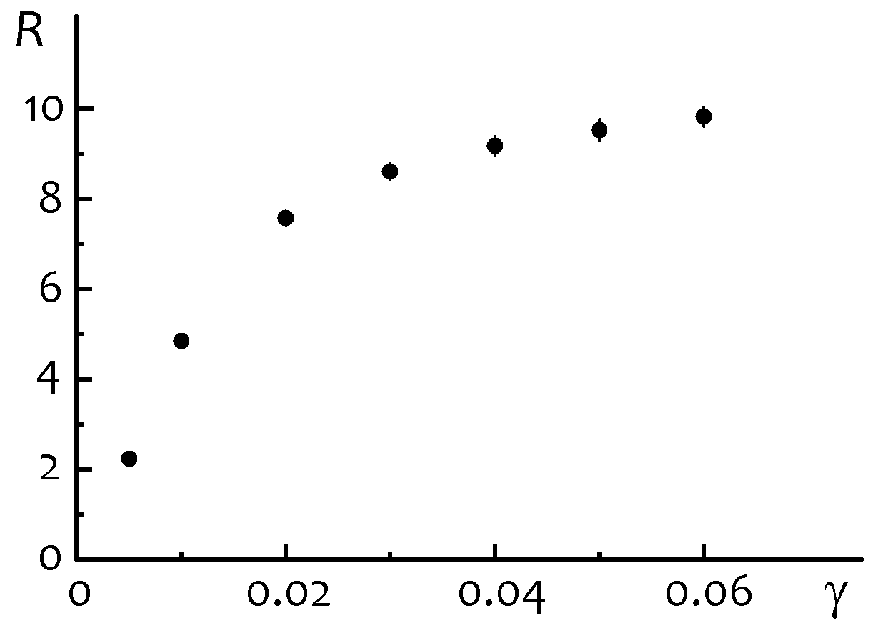}
\caption{Average cluster radius $R$ as a function of parameter $\gamma$. Simulation time is $t=5$.}
\label{fig6}
\end{figure}

Fig.~6 shows the dependence of average cluster radius $R$ on the parameter $\gamma$. It follows
from Fig.~6 that the strengthening of hydrophobic interactions at the interfacial boundary (increase
in the parameter $\gamma$) leads to an increase in the size of polymer aggregates.


\vskip20mm
\begin{center}
\bf 4. Conclusions
\end{center}

Using the methods of static and dynamic light scattering, dilute aqueous solutions of HPC (0.2~wt\%) with the admixtures
of Group I alkali metal chlorides (15.4~mmol/l) are studied. It is found that, at temperatures higher than the LCST,
there arises inverse scattering of light via the light reflection from the supramolecular structure of the researched
solutions. In the solutions with ions, there emerge large macromolecular associations (clusters) whose size is several
times larger than the wavelength of visible light. The temperature dependence of the intensity $J_{\text{R}}$ of backward
scattering is non-monotonic for specimens with salts; namely, as the temperature grows, the values of $J_{\text{R}}$
increase, reach a maximum, and finally decrease. The hydrodynamic diameter of clusters is demonstrated the same behavior
in corresponding solutions with ions, correlating with the $J_{\text{R}}$-values. There is a relation between the values
of $J_{\text{R}}$ and the sizes of polymer clusters, which was found in the framework of the tangent plane approximation
and Mie scattering theory. In particular, it is shown that the reduction of $J_{\text{R}}(T)$ in an interval of
$(55\div65)^{\circ}$C occurs due to the sedimentation of clusters rather than their decay, which was confirmed experimentally.

It is shown that in the vicinity of LCST the hydrophobic association of polymer molecules in solutions with salt ions
can be described in the framework of the Cahn-Hilliard model. It follows from the results of computer simulation that
the increase in the interfacial energy density at the polymer-solvent interface (strengthening of hydrophobic interactions)
leads to an increase in the size of polymer aggregates.

This study may provide useful information for the development of polymer films with a given structure and characteristics,
as well as new drug delivery systems.



\vspace{5mm}
\small
\begin{thebibliography}{100}

\bibitem{Okano1998}%
Okano,~T.
\textit{Biorelated Polymers and Gels: Controlled Release and Applications in Biomedical Engineering
(Polymers, Interfaces and Biomaterials)}, 1st ed.;
Academic Press: Cambridge, Massachusetts, USA, 1998;
ISBN 978-0125250900.


\bibitem{Kamide2005}%
Kamide,~K.
\textit{Cellulose and Cellulose Derivatives: Molecular Characterization and its Applications}, 1st ed.;
Elsevier Science: Amsterdam, Netherlands, 2005;
ISBN 978-0080454443.


\bibitem{Kabir2018}%
Kabir,~S.M.F.; Sikdar,~P.P.; Haque,~B.; Bhuiyan,~M.A.R.; Ali,~A.; Islam,~M.N.
Cellulose-based hydrogel materials: chemistry, properties and their prospective applications.
\textit{Prog. Biomater.} {\bf{2018}}, \textit{7}, 153–174.
[DOI: 10.1007/s40204-018-0095-0].


\bibitem{Tanaka2011}%
Tanaka,~F.
\textit{Polymer Physics: Applications to Molecular Association and Thermoreversible Gelation};
Cambridge University Press: Cambridge, UK, 2011.
[DOI: 10.1017/CBO9780511975691].


\bibitem{Bayer2012}%
Bayer, R.; Knarr, M.
Thermal precipitation or gelling behaviour of dissolved methylcellulose (MC)
derivatives—Behaviour in water and influence on the extrusion of ceramic pastes.
Part~1: Fundamentals of MC-derivatives.
\textit{J. Eur. Ceram. Soc.} {\bf{2012}}, \textit{32}, 1007-1018.
[DOI: 10.1016/j.jeurceramsoc.2011.1].


\bibitem{Tian2019}%
Tian,~Y.; Liu,~Y.; Ju,~B.; Ren,~X., Dai,~M.
Thermoresponsive 2-hydroxy-3-isopropoxypropyl hydroxyethyl cellulose
with tunable LCST for drug delivery.
\textit{RSC Adv.} {\bf{2019}}, \textit{9}, 2268–2276.
[DOI: 10.1039/c8ra09075k].


\bibitem{Gosecki2021}%
Gosecki,~M; Set\"al\"a,~H.; Virtanen,~T.; Ryan,~A.J.
A facile method to control the phase behavior of hydroxypropyl cellulose.
\textit{Carbohydr. Polym.} {\bf{2021}}, \textit{251}, 117015(7pp).
[DOI: 10.1016/j.carbpol.2020.117015].


\bibitem{Zhang2011}%
Zhang,~Z.; Chen,~L.; Zhao,~C.; Bai,~Y.; Deng,~M.; Shan,~H.; Zhuang,~X.; Chen,~X.; Jing,~X.
Thermo- and pH-responsive HPC-g-AA/AA hydrogels for controlled drug delivery applications.
\textit{Polymer} {\bf{2011}}, \textit{52}, 676–682.
[DOI: 10.1016/j.polymer.2010.12.048].


\bibitem{Bai2012}%
Bai,~Y.; Zhang,~Z.; Zhang,~A.; Chen,~L.; He,~C.; Zhuang,~X.; Chen,~X.
Novel thermo- and pH-responsive hydroxypropyl cellulose- and poly
(l-glutamic acid)-based microgels for oral insulin controlled release.
\textit{Carbohydr. Polym.} {\bf{2012}}, \textit{89}, 1207–1214.
[DOI: 10.1016/j.carbpol.2012.03.095].


\bibitem{Qiu2013}%
Qiu,~X.; Hu,~S.
``Smart'' Materials Based on Cellulose: A Review of the Preparations, Properties, and Applications.
\textit{Materials} {\bf{2013}}, \textit{6}, 738-781.
[DOI: 10.3390/ma6030738].


\bibitem{Ofridam2021}%
Ofridam,~F.; Tarhini,~M.; Lebaz,~N.; Gagni\`ere,~\'E.; Mangin,~D.; Elaissari,~A.
pH-sensitive polymers: Classification and some fine potential applications.
\textit{Polym. Adv. Technol.} {\bf{2021}}, \textit{32}, 1455–1484.
[DOI: 10.1002/pat.5230].


\bibitem{Xu2004-1}%
Xu,~Y.; Wang,~C.; Tam,~K.C.; Li,~L.
Salt-Assisted and Salt-Suppressed Sol-Gel Transitions of Methylcellulose in Water.
\textit{Langmuir} {\bf{2004}}, \textit{20}, 646–652.
[DOI: 10.1021/la0356295].


\bibitem{Xu2004-2}%
Xu,~Y.; Li,~L.; Zheng,~P.; Lam,~Y.C.; Hu,~X.
Controllable Gelation of Methylcellulose by a Salt Mixture.
\textit{Langmuir} {\bf{2004}}, \textit{20}, 6134–6138.
[DOI: 10.1021/la049907r].


\bibitem{Zheng2004}%
Zheng,~P.; Li,~L.; Hu,~X.; Zhao,~X.
Sol–Gel Transition of Methylcellulose in Phosphate Buffer Saline Solutions.
\textit{J. Polym. Sci. B Polym. Phys.} {\bf{2004}}, \textit{42}, 1849-1860.
[DOI: 10.1002/polb.20070].


\bibitem{Liu2008}%
Liu,~S.Q.; Joshi,~S.C.; Lam,~Y.C.
Effects of salts in the Hofmeister series and solvent isotopes on the gelation mechanisms
for hydroxypropylmethylcellulose hydrogels.
\textit{J. Appl. Polym. Sci.} {\bf{2008}}, \textit{109}, 363-372.
[DOI: 10.1002/app.28079].


\bibitem{Weibenborn2019}%
Wei{\ss}enborn,~E.; Braunschweig,~B.
Hydroxypropyl cellulose as a green polymer for thermo-responsive aqueous foams.
\textit{Soft Matter} {\bf{2019}}, \textit{15}, 2876-2883.
[DOI: 10.1039/c9sm00093c].


\bibitem{Nystrom1992}%
Nystr\"om,~B.; Roots,~J.; Carlsson,~A.; Lindman,~B.
Light scattering studies of the gelation process in an aqueous system of a non-ionic
polymer and a cationic surfactant.
\textit{Polymer} {\bf{1992}}, \textit{33}, 2875-2882.
[DOI: 10.1016/0032-3861(92)90071-4].


\bibitem{Joshi2011}%
Joshi,~S.C.
Sol-Gel Behavior of Hydroxypropyl Methylcellulose (HPMC) in Ionic Media Including Drug Release.
\textit{Materials} {\bf{2011}}, \textit{4}, 1861-1905.
[DOI: 10.3390/ma4101861].


\bibitem{Fettaka2011}%
Fettaka,~M.; Issaadi,~R.; Moulai-Mostefa,~N.; Dez,~I.; Le~Cerf,~D.; Picton,~L.
Thermo sensitive behavior of cellulose derivatives in dilute aqueous solutions:
From macroscopic to mesoscopic scale.
\textit{J. Colloid Interface Sci.} {\bf{2011}}, \textit{357}, 372-378.
[DOI: 10.1016/j.jcis.2011.02.041].


\bibitem{Almeida2014}%
Almeida,~N.; Rakesh,~L.; Zhao,~J.
Monovalent and divalent salt effects on thermogelation of aqueous hypromellose solutions.
\textit{Food Hydrocoll.} {\bf{2014}}, \textit{36}, 323-331.
[DOI: 10.1016/j.foodhyd.2013.10.020].


\bibitem{Lazarenko2022}%
Lazarenko M.; Nedilko S.; Gryn S.; Scherbatskyi V.; Kovalchuk V.; Lazarenko M.;
Sobchuk A.; Andrusenko D.; Alekseev O.
Influence of Na$^{+}$ and Cl$^{-}$ Ions on the Properties of Hydroxypropyl Cellulose Solutions.
In Proceedings of the 2022 IEEE 41st International Conference on Electronics and Nanotechnology
(ELNANO), Kyiv, Ukraine, October 10-14, 2022.
[DOI: 10.1109/ELNANO54667.2022.9927040].


\bibitem{Lazarenko2023}%
Lazarenko,~M.M.; Alekseev,~O.M.; Nedilko,~S.G.; Sobchuk,~A.O.; Kovalchuk,~V.I.;
Gryn,~S.V.;  Scherbatskyi,~V.P.; Tkachev,~S.Yu.; Andrusenko,~D.A.; Rudnikov,~E.G.;
Brytan,~A.V.; Yablochkova,~K.S.; Lysenkov,~E.A.; Dinzhos,~R.V.; Sabu,~T.; Abraham,~T.R.
Impact of the Alkali Metals Ions on the Dielectric Relaxation and Phase Transitions in
Water Solutions of the Hydroxypropylcellulose.
In \textit{NANO 2022: Nanoelectronics, Nanooptics, Nanochemistry and Nanobiotechnology, and Their Applications};
Fesenko,~O., Yatsenko,~L., Eds.;
Springer Proceedings in Physics: Cham, Switzerland, 2023;
Volume 297, pp.~37-68.
[DOI: 10.1007/978-3-031-42708-4].


\bibitem{Klement2007}%
Klement,~K.; Wieligmann,~K.; Meinhardt,~J.; Hortschansky,~P.; Richter,~W.; F\"andrich,~M.
Effect of Different Salt Ions on the Propensity of Aggregation and on the Structure of
Alzheimer’s $A\beta$(1-40) Amyloid Fibrils.
\textit{J. Mol. Biol.} {\bf{2007}}, \textit{373}, 1321-1333.
[DOI: 10.1016/j.jmb.2007.08.068].


\bibitem{Kirila2021}%
Kirila,~T.; Smirnova,~A.; Razina,~A.; Tenkovtsev,~A.; Filippov,~A.
Influence of Salt on the Self-Organization in Solutions of Star-Shaped Poly-2-alkyl-2-oxazoline
and Poly-2-alkyl-2-oxazine on Heating.
\textit{Polymers} {\bf{2021}}, \textit{13}, 1152(15pp).
[DOI: 10.3390/polym13071152].


\bibitem{Fritz2017}%
Fritz,~C.; Salas,~C.; Jameel,~H.; Rojas,~O.J.
Self-association and aggregation of kraft lignins via electrolyte and nonionic surfactant
regulation: stabilization of lignin particles and effects on filtration.
\textit{Nord. Pulp Paper Res. J.} {\bf{2017}}, \textit{32}, 572-585.
[DOI: 10.3183/npprj-2017-32-04\_p572-585\_rojas].


\bibitem{Wei2011}%
Wei,~C.-C.; Ge,~Z.-Q.
Influence of electrolyte and poloxamer 188 on the aggregation kinetics of solid
lipid nanoparticles (SLNs).
\textit{Drug Dev. Ind. Pharm.} {\bf{2011}}, \textit{38}, 1084–1089.
[DOI: 10.3109/03639045.2011.640331].


\bibitem{Pefferkorn1992}%
Pefferkorn,~E.
Electrolyte and Poly electrolyte Induced Aggregation of Colloids.
Mechanism of Colloid Destabilization.
\textit{Croat. Chem. Acta} {\bf{1992}}, \textit{65}, 309-326.
[https://hrcak.srce.hr/file/202172].


\bibitem{Hanus2001}%
Hanus,~L.H.; Hartzler,~R.U.; Wagner,~N.J.
Electrolyte-Induced Aggregation of Acrylic Latex. 1. Dilute Particle Concentrations.
\textit{Langmuir} {\bf{2001}}, \textit{17}, 3136–3147.
[DOI: 10.1021/la000927c].


\bibitem{Tavacoli2007}%
Tavacoli,~J.W.; Dowding,~P.J.; Routh,~A.F.
The polymer and salt induced aggregation of silica particles.
\textit{Colloids Surf. A Physicochem. Eng. Asp.} {\bf{2007}}, \textit{293}, 167–174.
[DOI: 10.1016/j.colsurfa.2006.07.025].


\bibitem{Khokhlov1997}%
Khokhlov,~A.R.; Dormidontova,~E.E.
Self-organization in ion-containing polymer systems.
\textit{Phys.-Uspekhi} {\bf{1997}}, \textit{40}, 109–124.
[DOI: 10.1070/PU1997v040n02ABEH000191].


\bibitem{Hussain2015}%
Hussain,~M.A.; Shah,~A.; Jantan,~I.; Shah,~M.R.; Tahir,~M.N.; Ahmad,~R.; Bukhari,~S.N.A.
Hydroxypropylcellulose as a novel green reservoir for the synthesis, stabilization,
and storage of silver nanoparticles.
\textit{Int. J. Nanomedicine} {\bf{2015}}, \textit{10}, 2079-2088.
[DOI: 10.2147/ijn.s75874].


\bibitem{Alharbi2023}%
Alharbi,~N.D.
Synthesis of composites from hydroxypropyl cellulose: iron (III) oxide nanoparticles.
\textit{Polym. Polym. Compos.} {\bf{2023}}, \textit{31}, 1012.
[DOI: 10.1177/09673911221149548].


\bibitem{Lee2020}%
Lee,~S.-J.; Kumar,~S.; Choi,~J.W.; Lee,~J.-S.
Coloration of colloidal polymer particles through selective extraction of
Mie backscattering for cation-responsible colorimetric sensors.
\textit{J. Colloid Interface Sci.} {\bf{2020}}, \textit{560}, 894-901.
[DOI: 10.1016/j.jcis.2019.10.073].


\bibitem{Connelly2017}%
Connelly,~K.; Wu,~Y.; Ma,~X.; Lei,~Y.
Transmittance and Reflectance Studies of Thermotropic Material for
a Novel Building Integrated Concentrating Photovoltaic (BICPV) `Smart' Window System.
\textit{Energies} {\bf{2017}}, \textit{10}, 1889(13pp).
[DOI: 10.3390/en10111889].


\bibitem{Nakamura2021}%
Nakamura,~A.; Ogai,~R.; Murakami,~K.
Development of smart window using an hydroxypropyl cellulose-acrylamide hydrogel
and evaluation of weathering resistance and heat shielding effect.
\textit{Sol. Energy Mater Sol. Cells} {\bf{2021}}, \textit{232}, 111348(8pp).
[DOI: 10.1016/j.solmat.2021.111348].


\bibitem{Harsh1991}%
Harsh,~D.C.; Gehrke,~S.H.
Controlling the swelling characteristics of temperature-sensitive cellulose ether hydrogels.
\textit{J. Control. Release} {\bf{1991}}, \textit{17}, 175-186.
[DOI: 10.1016/0168-3659(91)90057-K].


\bibitem{Narang2019}%
Narang,~A.S.; Badawy,~S.I.F.
\textit{Handbook of Pharmaceutical Wet Granulation: Theory and Practice in a Quality by Design Paradigm}, 1st ed.;
Academic Press: Cambridge, Massachusetts, USA, 2019;
ISBN 978-0128104606.


\bibitem{Kovalchuk2022}%
Kovalchuk, V.I.; Alekseev, O.M.; Lazarenko, M.M.
Turbidimetric Monitoring of Phase Separation in Aqueous Solutions of Thermoresponsive Polymers.
\textit{J. Nano- Electron. Phys.} {\bf{2022}}, \textit{14}, 01004(4pp).
[DOI: 10.21272/jnep.14(1).01004].


\bibitem{Kovalchuk2021}%
Kovalchuk, V.I.
Phase separation dynamics in aqueous solutions of thermoresponsive polymers.
\textit{Cond. Matt. Phys.} {\bf{2021}}, \textit{24}, 43601(9pp).
[DOI: 10.5488/CMP.24.43601].


\bibitem{Bass1979}%
Bass, F.G.; Fuks, I.M.
\textit{Wave Scattering from Statistically Rough Surfaces}, 1st ed.;
Pergamon Press: Oxford, UK, 1979;
ISBN 978-1483187754.


\bibitem{Lu2000}%
Lu, J.Q.; Hu, X.-H.; Dong, K.
Modeling of the rough-interface effect on a converging light beam propagating in a skin tissue phantom.
\textit{Appl. Opt.} {\bf{2000}}, \textit{39}, 5890-5897.
[DOI: 10.1364/ao.39.005890].


\bibitem{Lu1991}%
Lu, J.Q.; Maradudin, A.A.; Michel, T.
Enhanced backscattering from a rough dielectric film on a refecting substrate.
\textit{J. Opt. Soc. Am. B} {\bf{1991}}, \textit{8}, 311-317.
[DOI: 10.1364/JOSAB.8.000311].


\bibitem{Voronovich2007}%
Voronovich, A.G.
Tangent Plane Approximation and Some of Its Generalizations.
\textit{Acoust. Phys.} {\bf{2007}}, \textit{53}, 298–304.
[DOI: 10.1134/S1063771007030062].


\bibitem{Kodis1966}%
Kodis, R.
A Note on the Theory of Scattering from an Irregular Surface.
\textit{IEEE Trans. Antennas Propag.} {\bf{1966}}, \textit{14}, 77–82.
[DOI: 10.1109/TAP.1966.1138626].


\bibitem{Mie1908}%
Mie, G.
Beitr\"age zur Optik tr\"uber Medien, speziell kolloidaler Metall\"osungen.
\textit{Ann. Phys.} {\bf{1908}}, \textit{330}, 377–445.
[DOI: 10.1002/andp.19083300302].


\bibitem{Tzarouchis2018}%
Tzarouchis, D.; Sihvola, A.
Light Scattering by a Dielectric Sphere: Perspectives on the Mie Resonances.
\textit{Appl. Sci.} {\bf{2018}}, \textit{8}, 184–205.
[DOI: 10.3390/app8020184].


\bibitem{Kerker1983}%
Kerker, M.; Wang, D.S.; Giles, C.L.
Electromagnetic scattering by magnetic spheres.
\textit{J. Opt. Soc. Am.} {\bf{1983}}, \textit{73}, 765-767.
[DOI: 10.1364/JOSA.73.000765].


\bibitem{Matzler2002}%
M\"atzler, Ch.
\textit{MATLAB Functions for Mie Scattering and Absorption}, ver.~2;
Institute of Applied Physics: Bern, Switzerland, 2002;
Research Report No. 2002-11.


\bibitem{Mie2024}%
Mie Scattering Calculator.
Available online: https://omlc.org/calc/mie\_calc.html (accessed on January 20, 2024).


\bibitem{Maklakova2019}%
Maklakova, A.A.; Kulichikhin, V.G.; Malkin, A.Y.
The Formation and Elasticity of a Hydroxypropyl Cellulose Film at a Water-Air Interface.
\textit{Colloid J.} {\bf{2019}}, \textit{81}, 696-702.
[DOI: 10.1134/S1061933X19060103].


\bibitem{Li2015}%
Li, X.; Ji, G.; Zhang, H.
Phase transitions of macromolecular microsphere composite hydrogels based on the stochastic Cahn-Hilliard equation.
\textit{J. Comput. Phys.} {\bf{2015}}, \textit{283}, 81-97.
[DOI: 10.1016/j.jcp.2014.11.032].


\bibitem{DeGennes1980}%
De~Gennes, P.G.
Dynamics of fluctuations and spinodal decomposition in polymer blends.
\textit{J. Chem. Phys.} {\bf{1980}}, \textit{72}, 4756-4763.
[DOI: 10.1063/1.439809].


\bibitem{Flory1953}%
Flory, P.J.
\textit{Principles of Polymer Chemistry}, 1st ed.;
Cornell University Press, NY, USA, 1953;
ISBN 978-0801401343.


\bibitem{Cahn1958}%
Cahn, J.W.; Hilliard, J.E.
Free energy of a nonuniform system. I. Interfacial free energy.
\textit{J. Chem. Phys.} {\bf{1958}}, \textit{28}, 258-267.
[DOI: 10.1063/1.1744102].


\bibitem{Cahn1961}%
Cahn, J.
On spinodal decomposition.
\textit{Acta Metal.} {\bf{1961}}, \textit{9}, 795-801.
[DOI: 10.1016/0001-6160(61)90182-1].


\bibitem{Landau1980}%
Landau, L.D.; Lifshitz, E.M.
\textit{Course of Theoretical Physics, Vol. 5: Statistical Physics}, 3rd ed.;
Butterworth-Heinemann, Oxford, UK, 1980;
ISBN 978-0750633727.


\bibitem{Huggins2013}%
Huggins, M.L.
\textit{Physical Chemistry of High Polymers};
Literary Licensing LLC, Whitefish, USA, 2013;
ISBN 978-1258783365.


\bibitem{Lee2014}%
Lee D., Huh J.-Y., Jeong D., Shin J., Yun A., Kim J.
Physical, mathematical, and numerical derivations of the Cahn–Hilliard equation.
\textit{Comput. Mater. Sci.} {\bf{2014}}, \textit{81}, 216-225.
[DOI: 10.1016/j.commatsci.2013.08.027].


\bibitem{Orwoll2007}%
Orwoll~R.A., Arnold~P.A.
Polymer–Solvent Interaction Parameter $\chi$.
In: \textit{Physical Properties of Polymers Handbook};
Mark~J.E., Ed.;
Springer-Verlag, New York, 2007;
Chapter~14, pp.~233-257.
[ISBN: 978-0387312354].


\bibitem{HPC}%
Hydroxypropyl Cellulose.
Available online: https://www.chemsrc.com/en/cas/9004-64-2\_1198776.html
(accessed on March 24, 2025)


\bibitem{Bray1994}
Bray~A.J., Rutenberg~A.D.
Growth laws for phase ordering.
Phys. Rev. E, 1994, \textbf{49}, R27-R30.
[DOI: 10.1103/PhysRevE.49.R27].


\bibitem{Lifshitz1961}
Lifshitz~I.M., Slyozov~V.V.
The kinetics of precipitation from supersaturated solid solutions.
J. Phys. Chem. Solids, 1961, \textbf{19}, 35-50.
[DOI: 10.1016/0022-3697(61)90054-3].


\bibitem{Bray1993}
Bray~A.J.
Domain-growth scaling in systems with long-range interactions.
Phys. Rev. E, 1993, \textbf{47}, 3191-3195.
[DOI: 10.1103/PhysRevE.47.3191].









\end{thebibliography}
\end{document}